# Nonspecific protein-DNA binding is widespread in the yeast genome

Ariel Afek and David B. Lukatsky[1]
*Department of Chemistry, Ben-Gurion University of the Negev, Beer-Sheva 84105 Israel*

## Abstract

Recent genome-wide measurements of binding preferences of ~200 transcription regulators in the vicinity of transcription start sites in yeast, have provided a unique insight into the *cis*-regulatory code of a eukaryotic genome (Venters et al., *Mol. Cell* **41**, 480 (2011)). Here, we show that nonspecific transcription factor (TF)-DNA binding significantly influences binding preferences of the majority of transcription regulators in promoter regions of the yeast genome. We show that promoters of SAGA-dominated and TFIID-dominated genes can be statistically distinguished based on the landscape of nonspecific protein-DNA binding free energy. In particular, we predict that promoters of SAGA-dominated genes possess wider regions of reduced free energy compared to promoters of TFIID-dominated genes. We also show that specific and nonspecific TF-DNA binding are functionally linked and cooperatively influence gene expression in yeast. Our results suggest that nonspecific TF-DNA binding is intrinsically encoded into the yeast genome, and it may play a more important role in transcriptional regulation than previously thought.

[1] Corresponding author
Email: lukatsky@bgu.ac.il

# Introduction

## High-throughput measurements of protein-DNA binding *in vivo*

Specific transcription factor (TF) binding to genomic DNA in promoter regions is a key mechanism regulating gene expression in both prokaryotic and eukaryotic organisms. Recent advances in high-throughput methods of measuring TF-DNA binding preferences genome-wide *in vivo*, such as chromatin immunoprecipitation (ChIP) followed by microarray analysis (ChIP-chip), or followed by high-throughput sequencing analysis (ChIP-seq), provide a remarkable snapshot of the physical interaction map that exists within a living cell in different organisms (1-8). These measurements have demonstrated quite generally that TFs extensively bind thousands of active and inactive regions across the genome, and strikingly, in many cases no specific TF binding sites (TFBSs) can be identified in the regions of particularly strong binding (3-6). These observations have thus challenged the classical picture of specific TF-DNA binding. In their recent, seminal work, Venters et al., have measured binding preferences of 202 regulatory, DNA-binding proteins in three representative genomic regions in yeast (1). This work provides the most extensive view of TF-DNA binding in yeast up to date, and it concludes that over 90% of yeast promoter regions are significantly occupied by more than ten regulators, and ~10% are occupied by at least 75 regulators. The key, open question is what determines binding preferences of these regulators towards genomic DNA?

## Definition and design principles of nonspecific (non-consensus) protein-DNA binding

The existence and functional importance of nonspecific protein-DNA binding in *E. coli* were demonstrated in the early seventies of the last century in seminal experimental works of Riggs *et al.* (9), and Hinkle and Chamberlin (10); and in seminal theoretical works of von Hippel *et al.* (11-15), and Richter and Eigen (16). These early works suggested that DNA-binding proteins use different conformations in specific and nonspecific protein-DNA binding modes, respectively. Recent direct biophysical measurements performed both *in vivo* (17) and *in vitro* (18-22), unambiguously show that nonspecific protein-DNA binding is widespread in genomes of different organisms.

As presented in seminal works of von Hippel and Berg (13-15), the notion of nonspecific protein-DNA binding can be schematically described by two key, related mechanisms. The first mechanism is largely DNA sequence-independent, and it is entirely based (*i*) on the overall electrostatic attraction between DNA-binding proteins (such as TFs) and DNA, and (*ii*) on the overall geometry of DNA (13). The second mechanism assumes that for any sequence-specific DNA-binding protein, any DNA sequence, which is similar enough to canonical recognition motifs (consensus sequences) of this protein, possesses some residual protein-DNA binding affinity. For example, the yeast transcription factor Reb1 binds the TTACCCG motif with a relatively high affinity, and hence, any sequence similar to this consensus sequence is expected to possess a higher affinity to Reb1 than an entirely unrelated sequence (23). The fact that statistically, there is a high probability of having such sequence in many genomic locations by pure chance, might lead to nonspecific protein-DNA binding (13, 24).

We have recently suggested the existence of an additional, non-consensus nonspecific protein-DNA binding mechanism (25, 26). By using the term 'non-consensus nonspecific binding' we mean to express the fact that the predicted binding affinity is computed without experimental knowledge of the high-affinity sites for the TFs. In what follows, we always mean such non-consensus nonspecific TF-DNA binding. In particular, we predicted analytically that correlation properties of genomic DNA sequences generically regulate the nonspecific TF-DNA binding affinity (25). We use the term 'correlation' in order to describe statistically significant repeats of DNA sequence patterns. For example, we predicted that homo-oligonucleotide sequence correlations, where nucleotides of the same type are clustered together (such as poly(dA:dT) and poly(dC:dG) tracts) generically enhance the nonspecific TF-DNA binding affinity. Sequence correlations in which nucleotides of different types alternate have the opposite effect, reducing the nonspecific TF-DNA binding affinity (25). Due to the fact that the predicted effect stems from the intrinsic symmetry properties of DNA sequences, we suggested that it is quite general, and qualitatively robust with respect to microscopic details of the protein-DNA interaction potential (25). We also note that the predicted effect is entropy-dominated, and it assumes that TFs sample all possible binding sites along DNA (25, 26).

**Synopsis of obtained results**

Here, having obtained experimental binding preferences of 202 DNA binding proteins (1), we thought to answer the question what role does nonspecific (non-consensus) protein-DNA binding play in a living yeast cell, genome-wide?

In order to address this question, here we compute the nonspecific binding free energy of random protein-DNA binders. We use the term 'random binder' in order to emphasize the fact that model TFs bind genomic DNA nonspecifically. We compute statistical properties of such nonspecific binding. Strikingly, we show that nonspecific binding alone can explain statistical binding preferences observed experimentally. Our results provide further support of the hypothesis that nucleosome occupancy in yeast is significantly influenced by nonspecific TF-DNA binding (26).

We note that in the experiments that we are using for this analysis (1), TFs can be cross-linked and immunoprecipitated in association with a given DNA segment by virtue of at least four kinds of interactions. (*i*) Binding to the local DNA. (*ii*) Cooperative binding to a combination of local DNA and other locally-bound TFs. (*iii*) Cooperative binding only to other locally-bound TFs and not to DNA. (*iv*) Binding to nascent RNA transcripts and/or proteins bound to nascent RNA transcripts. Our theoretical analysis of protein-DNA binding affinity focuses largely on mechanism (*i*). Yet, due to the fact that all our predictions are statistical in nature, and the number of experimentally measured TFBSs is very large, we suggest that all our conclusions are quite general, and most likely represent the statistical law, rather than the exception.

This paper is organized as follows. First, we describe our method to compute the nonspecific TF-DNA binding free energy landscape. Second, we show that nonspecific TF-DNA binding significantly influences experimentally observed TF-DNA binding preferences in promoter regions of the yeast genome, **Figure 1** and **Figure 2**. Third, we show that promoter regions of highly regulated (e.g. SAGA-dominated) and weakly regulated (e.g. TFIID-dominated) genes are characterized by distinct profiles of the nonspecific binding free energy, **Figure 3**. In conclusion, we show that the level of gene expression in yeast grown in YPD medium is correlated with the landscape of the nonspecific binding free energy in promoter regions, **Figure 4** and **Figure 5**.

# Results

## Model free energy of nonspecific TF-DNA binding

We begin by computing the free energy of nonspecific TF-DNA binding in three genomic locations surrounding the transcription start sites of 4962 highly confident yeast transcripts (Materials and Methods). We use the following terms, adopted from ref. (1), to describe these three types of locations: transcription start sites (TSSs) located in the interval (-90,-30), upstream activating sequences (UASs) located in the interval (-320,-260), and open reading frames (ORFs) located in the 3' end of the coding regions of genes (Materials and Methods). The occupancy of 202 transcription regulators (we use the term, TFs, to describe the regulators) where experimentally determined in these three locations in ref. (1). We note that we use a conventional abbreviation, 'TSS', to describe both the transcription start site, where zero of our coordinate system is positioned in each gene, and the region in the upstream vicinity of the TSS site, located in the interval (-90,-30), as defined in ref. (1). This coincidence should not lead to confusion, as the precise meaning of 'TSS' will be clear from the context in each case.

In order to compute the free energy of nonspecific TF-DNA binding in each genomic location specified above, we use a simple variant of the Berg-von Hippel model (14, 15), developed recently (25, 26). In particular, we can assign the free energy of nonspecific TF-DNA binding to each DNA base pair along the genome in the following way. First, we position a midpoint of the sliding window of width $L = 50$ bp at a given genomic coordinate.

Second, we compute the partition function of the model TF sliding along the sliding window:

$$Z = \sum_{i=1}^{L} \exp(-U(i) / k_B T), \qquad \text{Eq. (1)}$$

where $k_B$ is the Boltzmann constant, $T$ is the temperature, and $U(i)$ is the TF-DNA binding energy at the position $i$ within the sliding window. The TF-DNA binding energy of the TF forming $M$ contacts with DNA basepairs, at a given position $i$ within the sliding window:

$$U(i) = -\sum_{j=i}^{M+i-1} \sum_{\alpha=1}^{4} K_\alpha s_\alpha(j), \qquad \text{Eq. (2)}$$

where $s_\alpha(j)$ is a four-component vector of the type $(\delta_{\alpha A}, \delta_{\alpha T}, \delta_{\alpha C}, \delta_{\alpha G})$, specifying the identity of the base-pair at each DNA position $j$, with $\delta_{\alpha\beta} = 1$, if $\alpha = \beta$, and $\delta_{\alpha\beta} = 0$, if $\alpha \neq \beta$. For example, if a given DNA site, $j$, is occupied by the A nucleotide, this vector takes the form: $(1,0,0,0)$; if the site $j$ is occupied by the C nucleotide, this vector is $(0,0,1,0)$. Within the framework of our model, each TF is fully described by four energy parameters, $K_A$, $K_T$, $K_C$, and $K_G$ (25). In order to model nonspecific TF-DNA binding, we generate an ensemble of 250 TFs, and for each TF we draw the energies $K_A$, $K_T$, $K_C$, and $K_G$ from the Gaussian probability distributions, $P(K_\alpha)$, with zero mean and standard deviations, $\sigma_\alpha = 2k_B T$, where $\alpha = \mathrm{A, T, C, G}$. Therefore, each random realization of $P(K_\alpha)$ describes one TF.

Third, we compute the free energy of nonspecific TF-DNA binding, $F = -k_B T \ln Z$, for each randomly generated TF in this sliding window. We always consider the difference, $\Delta F = F - F_\infty$, where $F_\infty$ is the free energy computed for randomized sequence of the same width, $L$, and averaged over 50 random realizations of this sequence, for a given TF. This normalization procedure removes the effect of the compositional bias, and allows us to compare the free energies of nonspecific TF-DNA binding in different genomic regions, despite the variation of the average nucleotide composition along the genome. We perform this calculation for all 250 randomly generated TFs. We note that the results are very weakly dependent of the sliding window width, $L$ (data not shown).

Fourth, we move the sliding window along the genome, assigning the free energy of nonspecific TF-DNA binding for each randomly generated TF, to each genomic coordinate in steps of 4 bp, within the three regions described above: TSSs, UASs, and ORFs, respectively. This procedure allows us to perform a direct comparison of the TF occupancy in these genomic regions between the model and experiment (1).

**Nonspecific binding significantly influences experimentally observed TF-DNA binding preferences**

We now seek to answer the question to what extent does nonspecific TF-DNA binding influence experimentally observed TF binding preferences within the TSS, UAS, and ORF regions, respectively? In order to answer this question, we first select 10% highest and 10% lowest average TF occupancy genes (see Materials and Methods for the definition of the experimentally measured, average TF occupancy). We perform such selection separately with respect to TF occupancy in the TSS, UAS, and ORF regions, respectively.

Next, we compute the profile of nonspecific TF-DNA binding free energy, within the range $(-384,384)$, for the highest and the lowest TF occupancy genes, selected in both the TSS regions, **Figure 1 A,** and UAS regions, **Figure 1 B**. For each gene, we compute the free energy (normalized per bp), averaged with respect to 250 model TFs, $\Delta f = \langle \Delta F \rangle_{TF} / M$ . After that, we compute the average of $\Delta f$ with respect to the selected 10% highest and 10% lowest average TF occupancy genes, aligned with respect to their transcription start sites, $\langle \Delta f \rangle = \langle \langle \Delta F \rangle_{TF} \rangle_{seq} / M$ , where the second average, $\langle ... \rangle_{seq}$, describes the averaging with respect to the aligned sequences. In both the TSS and UAS regions we observe that the highest TF occupancy genes exhibit a lower free energy of nonspecific TF-DNA binding compared with the lowest TF occupancy genes that exhibit a higher free energy. This result is statistically significant with the *p*-values, $p \simeq 0.01$ and $p \simeq 0.007$, for the TSS and UAS regions, respectively (Materials and Methods). A different definition of the average TF occupancy leads to similar results, **Figure S1**.

The free energy of nonspecific TF-DNA binding significantly correlates with the experimentally observed average TF occupancy within the entire dynamic range of the occupancy values in both TSS and UAS regions, **Figure 1 C** and **D**. Here, we ordered genes in bins with respect to the value of their average TF occupancy, and computed the minimal free energy, $\min(\Delta f)$, for each sequence, in each bin. It is remarkable that in both TSS and UAS regions the linear fits exhibit identical slopes, **Figure 1 C** and **D**. We conclude therefore that nonspecific TF-DNA binding significantly influences binding preferences in promoter regions of the majority of transcription regulators in yeast.

We note that theoretical analysis of the nonspecific TF-DNA binding free energy in the ORF regions does not show statistically significant correlation with the experimentally measured average TF occupancy in these regions, unlike the trend described above in the TSS and UAS regions (data not shown). Overall, the magnitude of the nonspecific TF-DNA binding free energy in the ORFs regions is weak compared to the TSS and UAS regions, as **Figure 1** clearly demonstrates.

We now demonstrate that the experimentally measured *cumulative* TF occupancy in the promoter regions as compared to coding regions (1), is also accurately predicted within the framework of our model. In particular, in order to define the cumulative TF occupancy theoretically, we assume in the computational procedure that if the minimal binding free energy, $\min(\Delta F)$, of a given TF within a given genomic region for a particular gene is less than a certain cutoff value, then it binds to this region. **Figure 2 A** and **B** show the result of such comparison for promoter regions (combined binding to TSSs and UASs) and coding regions (ORFs), based on the theoretical calculation (**Figure 2 A**) and experimental measurements (**Figure 2 B**). These results are highly statistically significant, as the biological error bars demonstrate, **Figure 2 A** and **B**. The agreement between the theory and experiment holds quantitatively significant for a wide, physically relevant range of the free energy cutoff values (data not shown). Notably, when we compute the cumulative TF occupancy, separating promoter regions into TSSs and UASs, we observe a disagreement with the experimental data for TSSs and UASs. In particular, our model predicts that TSS regions possess a higher propensity for nonspecific TF-DNA binding than UAS regions, while the experimental data show an opposite trend. The reason for this disagreement is currently not understood. At least two additional factors may be responsible for the observed discrepancy. First, as we mentioned in the introduction, several types of interactions determine the measured TF occupancy *in vivo* (1). These are (*i*) direct binding to the local DNA; (*ii*) cooperative binding to a combination of local DNA and other locally-bound TFs; (*iii*) cooperative binding only to other locally-bound TFs and not to DNA; and (*iv*) binding to nascent RNA transcripts and/or proteins bound to nascent RNA transcripts. Our current model takes into account only mechanism (*i*). Second, our theoretical approach is purely equilibrium, while kinetic barriers might significantly influence TF binding preferences *in vivo* (17).

We conclude, therefore, that nonspecific TF-DNA binding alone can accurately account for the experimentally observed differences in the cumulative TF occupancy of promoter regions as compared with coding regions, yet our model fails to predict the experimentally measured, absolute differences between TSSs and UASs within the promoter regions (1).

## Discussion and conclusion

### Nonspecific binding distinguishes between SAGA-dominated and TFIID-dominated genes

Genome-wide studies found that ~90% of the yeast genome is TFIID dominated, while the remaining ~10% of genes are SAGA dominated (1, 27, 28). SAGA-dominated genes typically contain the TATA box, and they are highly regulated compared to TFIID-dominated genes, which are typically TATA-less (27, 28). The majority of the known stress-response genes in yeast tend to belong to the SAGA-dominated class (27, 28). It is also known that the high transcriptional plasticity genes are enriched in SAGA-dominated genes compared to the low transcriptional plasticity genes (29, 30). It was concluded in a recent study by Venters et al. (1) that SAGA-dominated/TATA-containing genes were occupied by a larger variety of regulators compared to TFIID-dominated/TATA-less genes. Here, we show that the nonspecific TF-DNA binding free energy can qualitatively explain the observed difference in the TF occupancy between these two classes of genes.

In particular, we computed the profile of nonspecific TF-DNA binding free energy for 40 highly confident SAGA-dominated (TATA-containing) genes, and 178 TFIID-dominated (TATA-less, non-ribosomal protein) genes, respectively (1). The key conclusion here is that SAGA-dominated genes exhibit a wider region of the reduced free energy (within the interval (-384,384) around the TSS site) compared to TFIID-dominated genes, **Figure 3 A**, with the *p*-value, $p \simeq 4.4 \times 10^{-4}$. We suggest, therefore, that the reduced free energy of nonspecific TF-DNA binding plays the role of an effective, attractive potential that facilitates nonspecific binding to promoters of both SAGA-dominated and TFIID-dominated genes, however, the predicted effect is stronger for the former group, leading to a higher average TF occupancy of SAGA-dominated genes. In order to test the functional robustness of our conclusion, we selected two larger groups of 15% highest and 15% lowest transcriptional plasticity genes, respectively (732 genes in each group) (29, 30) (Materials and Methods). The high-plasticity genes are enriched in SAGA-dominated genes (272 SAGA genes out of 723 high-plasticity genes) compared to the low-plasticity genes (3 SAGA genes out of 732 low-plasticity genes), with $p < 10^{-6}$. The free energy calculation performed for these two groups shows that the high-plasticity genes possess a wider region of the reduced free energy compared to the low-plasticity genes, **Figure 3 B**. Therefore we conclude quite generally, that nonspecific TF-DNA binding significantly influences functional properties of yeast genes, and presumably, it facilitates the search of specific TF binding sites in promoter regions. Promoters of highly regulated genes appear to possess a wider region of the reduced nonspecific TF-DNA binding free energy (on average), compared to weakly regulated genes.

**Gene expression is correlated with nonspecific TF-DNA free energy landscape**

We ask now the question: How is the gene expression in yeast influenced by nonspecific TF-DNA binding? In order to answer this question, first, for each gene we computed the average free energy profile, $\Delta f = \langle \Delta F \rangle_{TF} / M$, in the promoter region within the interval (-150,0). Second, for each gene we found the minimum of $\Delta f$ within this interval, $\Delta f_{\min} = \min(\Delta f)$. As a result, we observe a statistically significant correlation of $\Delta f_{\min}$ with the level of gene expression (31), **Figure 4 A**. This result suggests that nonspecific TF-DNA binding influences gene expression in yeast. In order to obtain a deeper insight into a relationship between nonspecific TF-DNA binding and gene expression, we introduce the notion of nonspecific transcription factor binding nucleotides (TFBNs). In particular, we define a given position within the genome as being the nonspecific TFBN, if the computed average free energy of nonspecific TF-DNA binding in this genomic location is less than a certain cutoff value. The correlation between the number of nonspecific TFBNs within the interval $(-150,0)$ and the gene expression level is shown in **Figure 4 B**. Statistically significant correlation persists for a wide range of the cutoff values (data not shown). In order to understand how specific and nonspecific TF-DNA binding is related, as far as gene expression is concerned, we also present the correlation between the number of specific TFBSs and the level of gene expression, **Figure 4 C** and **D**, where the information about specific TFBSs is extracted from ref. (32). We conclude, therefore, that first, the propensity of promoter regions towards nonspecific TF binding statistically significantly influences gene expression, and second, specific and nonspecific binding are functionally linked.

Next, we seek to understand whether the obtained relationship between nonspecific TF-DNA binding within the interval $(-150,0)$ and the level of gene expression persists within the TSS, UAS, and ORF regions, respectively. In order to answer this question, we first present the correlation between the experimentally measured average TF occupancy in each region and the level of gene expression, **Figure 5 A**, **B**, and **C**. Remarkably, a statistically significant correlation is observed in all three regions. **Figure 5 D**, **E**, and **F** present the correlation between the computed minimal free energy of nonspecific binding and the level of gene expression, within each region, TSS, UAS, and ORF, respectively. The strongest correlation is observed for the TSS regions, the correlation in the UAS regions is also significant, but contrary to experimental results, we do not observe correlation between $\Delta f_{\min}$ and the level of gene expression in the ORF regions, **Figure 5 F**. We conclude, therefore, that in yeast, the strength of nonspecific TF-DNA binding is encoded and fine-tuned within a wide interval (of ~300 bp) in promoter regions, and it influences the level of gene expression. Our results suggest that in the coding regions the effect of nonspecific TF-DNA binding on gene expression is insignificant, and it is likely that other factors, such as specific ATP-dependent chromatin modifying factors, might play a dominant role there.

**Conclusion**

In summary, here we showed, first, that nonspecific TF-DNA binding significantly influences binding preferences of ~200 transcription regulators in promoter regions of the yeast genome. Second, our analysis suggests that specific and nonspecific binding are functionally linked. Third, we observed quite generally, that promoter regions of highly regulated genes, such as SAGA-dominated genes, possess a wider region of the reduced nonspecific binding free energy compared to promoter regions of weakly regulated genes, such as TFIID-dominated genes. This qualitatively explains the experimental observation in ref. (1) that promoters of SAGA-dominated genes are more highly occupied (on average) than promoters of TFIID-dominated genes. Fourth, we showed that the landscape of nonspecific binding free energy in promoter regions correlates with the level of gene expression.

We emphasize that in order to compute the nonspecific TF-DNA binding free energy genome-wide, we used a highly simplified biophysical model. Despite the simplicity of this model, we suggest that our conclusions are quite general, and most likely they represent the statistical rule, rather than the exception. The generality of our conclusions stems from the fact that the computed, location-dependent affinity of the genome for nonspecific TF-DNA binding is dominated exclusively by the symmetry of DNA sequence correlations, and this affinity is expected to be weakly dependent of microscopic details of the model.

## Materials and Methods

**Gene set**

In our analysis we used a highly confident set of 4962 yeast genes from ref. (33). We used the following terms, adopted from ref. (1), to describe three types of genomic regions: transcription start sites (TSSs) located in the interval (-90,-30), upstream activating sequences (UASs) located in the interval (-320,-260), and open reading frames (ORFs) located in the 3' end of the coding regions. The zero of the coordinate system is located in the transcription start site for each gene. We note that in ref. (1) the ORF regions were positioned in slightly different genomic locations in different genes, downstream of the transcription start site, Table S2 of ref. (1). In our analysis we used the precise, experimental location of the ORF for each gene.

**Experimental TF occupancy**

The experimental average TF occupancy in each of the three genomic locations, TSSs, UASs, and ORFs, measured in ref. (1), is defined for each gene in the following way. For each gene, in each genomic location, we compute the average occupancy of all regulators from Table S2 of ref. (1), measured at temperature of 25C. Only regulators with the occupancy above 5% threshold for false discover rate (FDR) reported in ref. (1) are taken into account in the calculation of the average TF occupancy. At the end of this procedure, each genomic location, TSSs, UASs, and ORFs, respectively, in each gene is assigned a value of the average TF occupancy.

**Gene expression data**

The experimental gene expression data in YPD medium is taken from ref. (31).

**SAGA-dominated and TFIID-dominated genes**

In order to compute **Figure 3 A**, we extracted the sets of 40 SAGA-dominated, TATA-containing genes and 178 TFIID-dominated, TATA-less, non-ribosomal protein genes, from Table S6 of ref. (1). The extended list of all known SAGA-dominated and TFIID-dominated genes is taken from ref. (28).

**Transcriptional plasticity**

In order to compute **Figure 3 B**, we used the classification of transcriptional plasticity from ref. (29), and refined in ref. (30).

***p*-value calculations**

**Figure 1 A and B:** In order to compute the *p*-values, first, we selected $10^5$ pairs of groups of randomly chosen 496 genes. Each pair of groups represents randomized analogs of the highest occupancy and the lowest occupancy genes, respectively. Second, for each of these pairs of random groups we computed the free energy of nonspecific binding, as described above. Third, within each region of interest (TSS or UAS), we computed the difference between the minima of the average free energy of nonspecific binding, $\langle \Delta f \rangle_{\min}$, for the corresponding pairs of groups. Finally, we computed the probability that this difference is equal or larger than the actual value of the difference. The latter probability was taken as the *p*-value.

**Figure 3 A:** In order to compute the *p*-value, we first compiled $3 \times 10^5$ pairs of groups of randomly chosen 178 and 40 genes, respectively. These groups represent the randomized analogs for TFIID and SAGA genes, respectively. Second, for each of these pairs of random groups we computed the average free energies, $\langle \Delta f \rangle$, of nonspecific binding separately for randomized TFIID and randomized SAGA groups, as described above. Third, for each pair of randomized groups we computed the difference of the integrated free energy within the interval $(-384,100)$ between randomized TFIID and SAGA groups. Finally, we computed the probability that this difference is equal or larger than the actual value of the difference. The latter probability was taken as the *p*-value.

## Acknowledgements

We thank Itay Tirosh for providing us the data on transcriptional plasticity. We also thank the anonymous referee who helped to improve our paper. D.B.L. acknowledges the financial support from the Israel Science Foundation (ISF) grant 1014/09. A.A. is a recipient of the Lewiner graduate fellowship.

## Supporting Material

One supporting figure.

## Figure Legends

**Figure 1. A.** Average free energy of nonspecific TF-DNA binding per bp, $\langle \Delta f \rangle = \langle \langle \Delta F \rangle_{TF} \rangle_{seq} / M$, computed within the interval (-384,384) for the two groups of genes selected according to the experimentally measured average TF occupancy in the TSS region: 10% highest TF occupancy in the TSS region (red) and 10% lowest TF occupancy in the TSS region (blue). Each group contains 496 genes. Horizontal bar, marked 'TSS', on the *x*-axis, shows the corresponding region where the TF occupancy was measured. **B.** Similar to **(A)**, but the two groups of genes are selected according to the experimentally measured average TF occupancy in the UAS region. Horizontal bar, marked 'UAS', on the *x*-axis, shows the corresponding region where the TF occupancy was measured. **C.** Correlation between the minimal value of the free energy of nonspecific TF-DNA binding within the TSS regions, $\Delta f_{\min} = \min(\langle \Delta F \rangle_{TF}) / M$, and the average TF occupancy within this region. Genes were binned into ten bins according to the value of the average TF occupancy. Each point in the graph corresponds to the average, $\langle \Delta f_{\min} \rangle$, for the genes in a given bin plotted as a function of the experimentally measured average TF occupancy for the genes in this bin. **D.** Analogous to **(C)**, but for $\langle \Delta f_{\min} \rangle$ computed within the UAS regions, plotted versus the average TF occupancy measured within the UAS regions, as described in **(C)**.

**Figure 2. A and B.** Number of promoter regions (TSSs and UASs) (black) and coding regions (ORFs) (red) occupied by the number of regulators (i.e. TFs) indicated along the *x*-axis, as computed using the model of nonspecific TF-DNA binding (**A**) and experimental data from ref. (1) (**B**). This corresponds to Figure 2A of ref. (1). In the computational prediction we assumed that a given genomic region is occupied by a given TF if the minimal free energy of nonspecific TF-DNA binding (within this genomic region) is less than the cutoff value of $-1\, k_B T$, and we used 250 TFs in the computation (Materials and Methods). In order to compute error bars, we divided all genes into four sub-groups, and computed the corresponding occupancy separately for each sub-group. The error bars are defined as one standard deviation of the occupancy between the sub-groups. Inset in each panel shows the occupancy for the entire set of ~5000 genes. **C** and **D**. Analogous to the insets in **A** and **B**, but with the cumulative TF occupancy computed separately for TSSs and UASs. We used $M = 8$ for the TF length in all our calculation of the free energy.

**Figure 3. A.** Average free energy of nonspecific TF-DNA binding per bp, $\langle \Delta f \rangle$, computed within the interval (-384,384) for the highly confident SAGA-dominated and TFIID-dominated groups of genes, respectively. There are 40 SAGA-dominated, TATA-containing genes and 178 TFIID-dominated, TATA-less, non-ribosomal protein genes, respectively (these highly-confident groups are taken from ref. (1)). **B.** Average free energy of nonspecific TF-DNA binding per bp, $\langle \Delta f \rangle$, computed within the interval (-384,384) for the high and low transcriptional plasticity genes, respectively. There are 732 genes in each group. In order to compute error bars, we divided each group of genes into five arbitrary subgroups, computed $\langle \Delta f \rangle$ in each of the subgroups, and computed the standard deviation of $\langle \Delta f \rangle$ between the subgroups. Error bars correspond to one standard deviation.

**Figure 4. A.** Correlation between the minimal value of the free energy of nonspecific TF-DNA binding in the promoter region, within the interval (-150,0), $\Delta f_{\min} = \Delta F_{\min} / M$, and the average

value of gene expression within this region. All ~5000 genes were binned into 25 bins according to the level of gene expression. Each point in the graph corresponds to the average, $\langle \Delta f_{\min} \rangle$, for the genes in a given bin plotted as a function of the experimentally measured average level of gene expression for the genes in this bin. **B.** Correlation between the computed number of nonspecific TF binding nucleotides (TFBNs) within the interval (-150,0), and the level of gene expression. A given genomic coordinate is assigned to belong to nonspecific TFBN, if the average free energy of nonspecific TF-DNA binding per nucleotide is smaller than a given cutoff value, $\Delta f < -0.25\ k_B T$. **C.** Correlation between the number of specific TFBSs and the gene expression. The information about specific TFBSs is taken from ref. (32). **D.** Correlation between the number of specific TFBSs and the number nonspecific TFBNs. The binning in (**B**), (**C**), and (**D**) is preformed as in (**A**).

**Figure 5.** Analysis of experimental results from ref. (1): Correlation between the average TF occupancy and the level of gene expression for TSS (**A**), UAS (**B**), and ORF (**C**) regions, respectively. Genes were binned into 25 bins according to the level of gene expression. Each point in the graph corresponds to the average, experimental TF occupancy for the genes in a given bin plotted as a function of the experimentally measured, average level of gene expression for the genes in this bin. Correlation between the computed, average value of the minimal free energy of nonspecific TF-DNA binding, $\langle \Delta f_{\min} \rangle$, and the level of gene expression for TSS (**D**), UAS (**E**), and ORF (**F**) regions, respectively. The binning is performed as explained above.

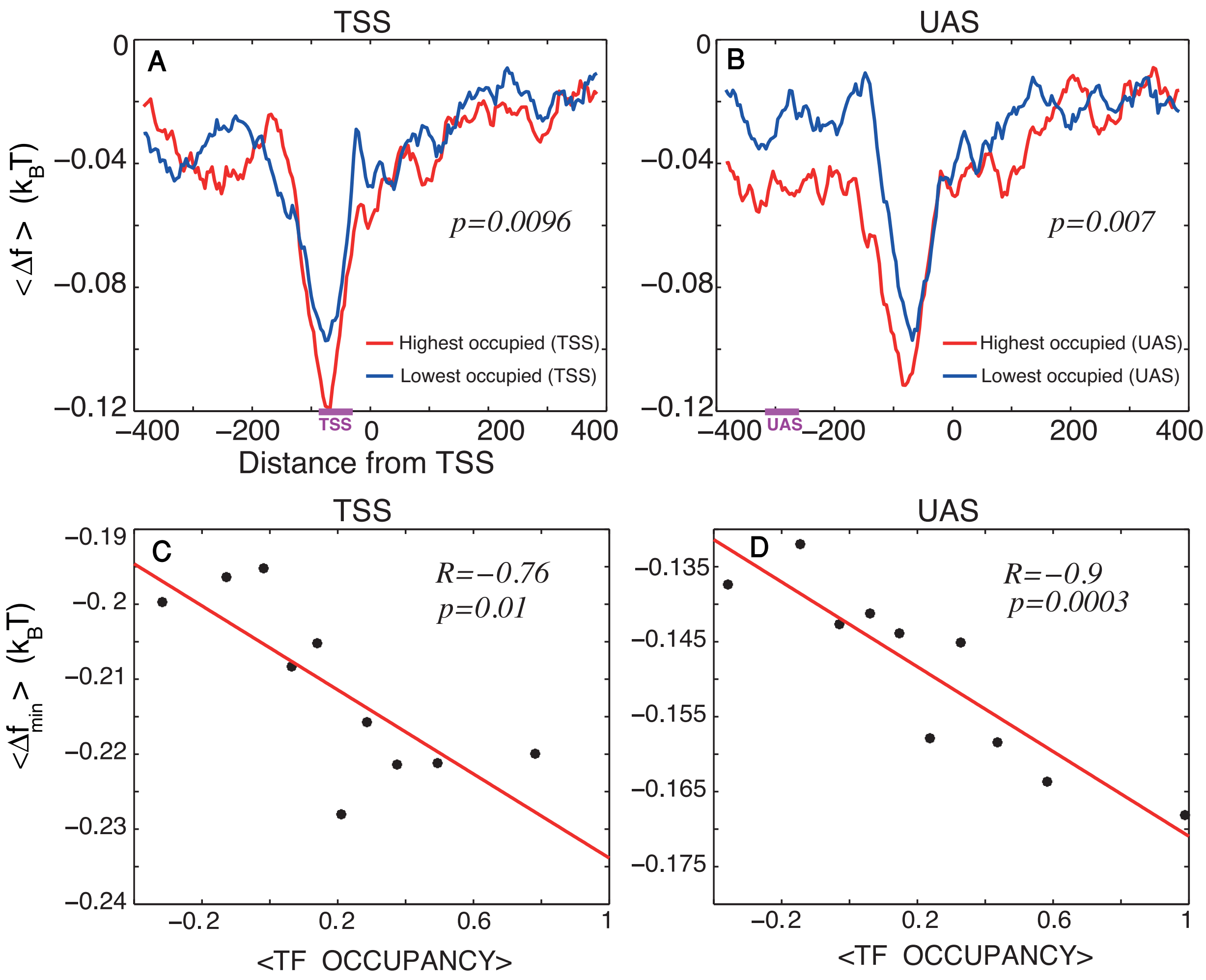

**Figure 1**

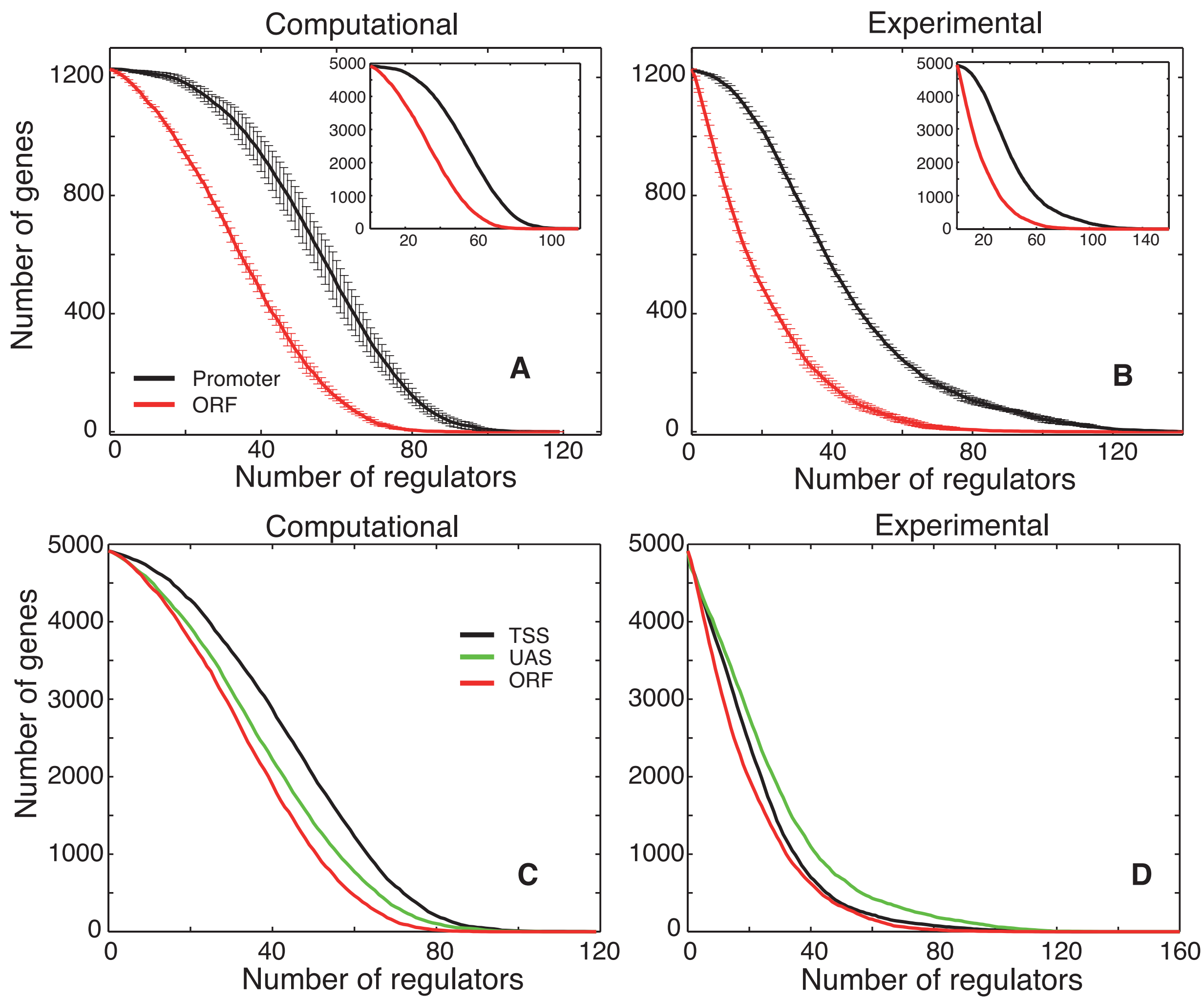

**Figure 2**

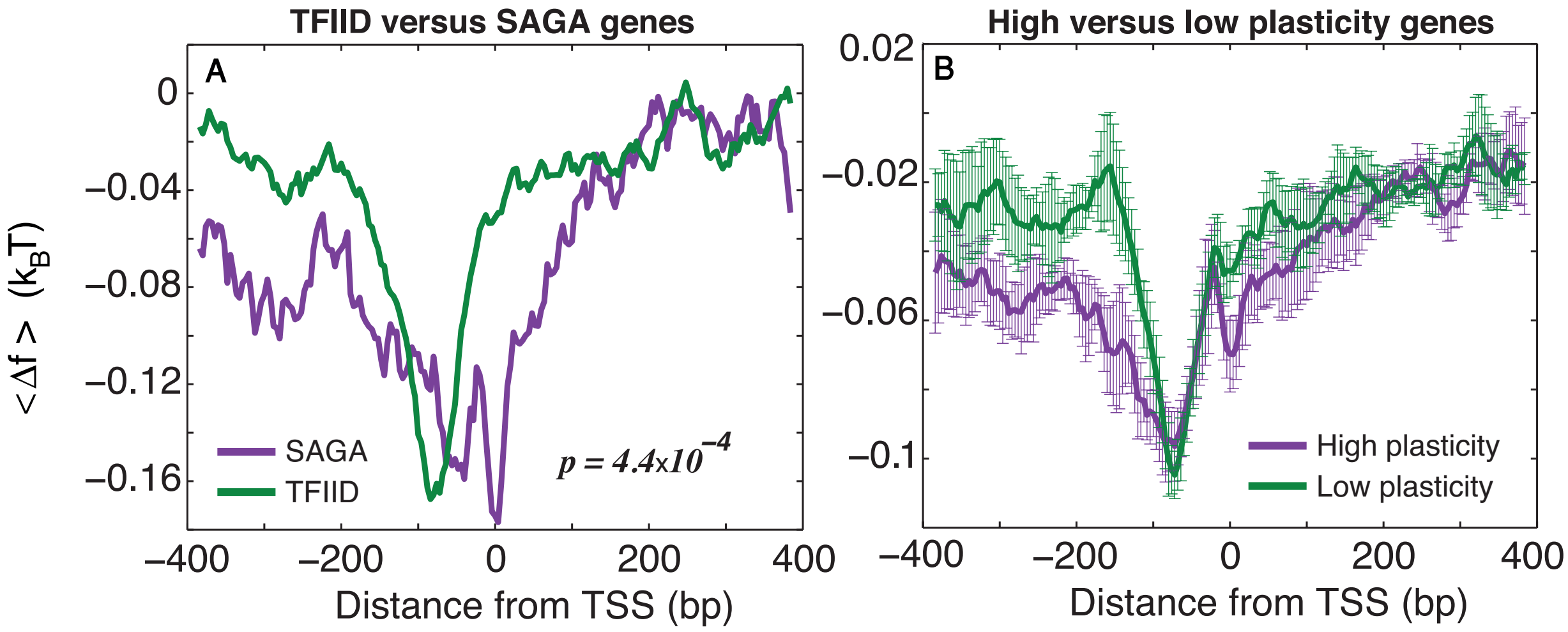

**Figure 3**

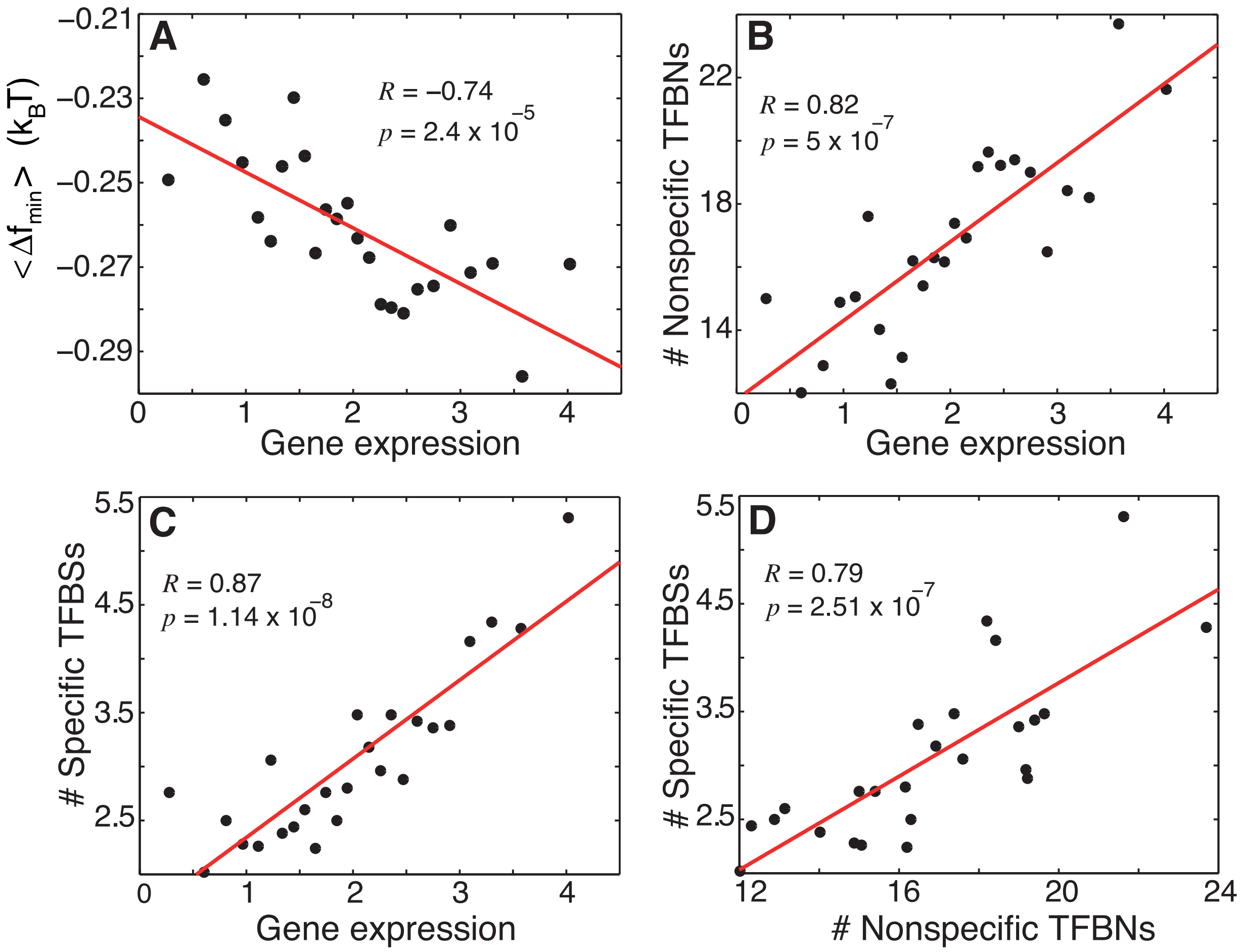

**Figure 4**

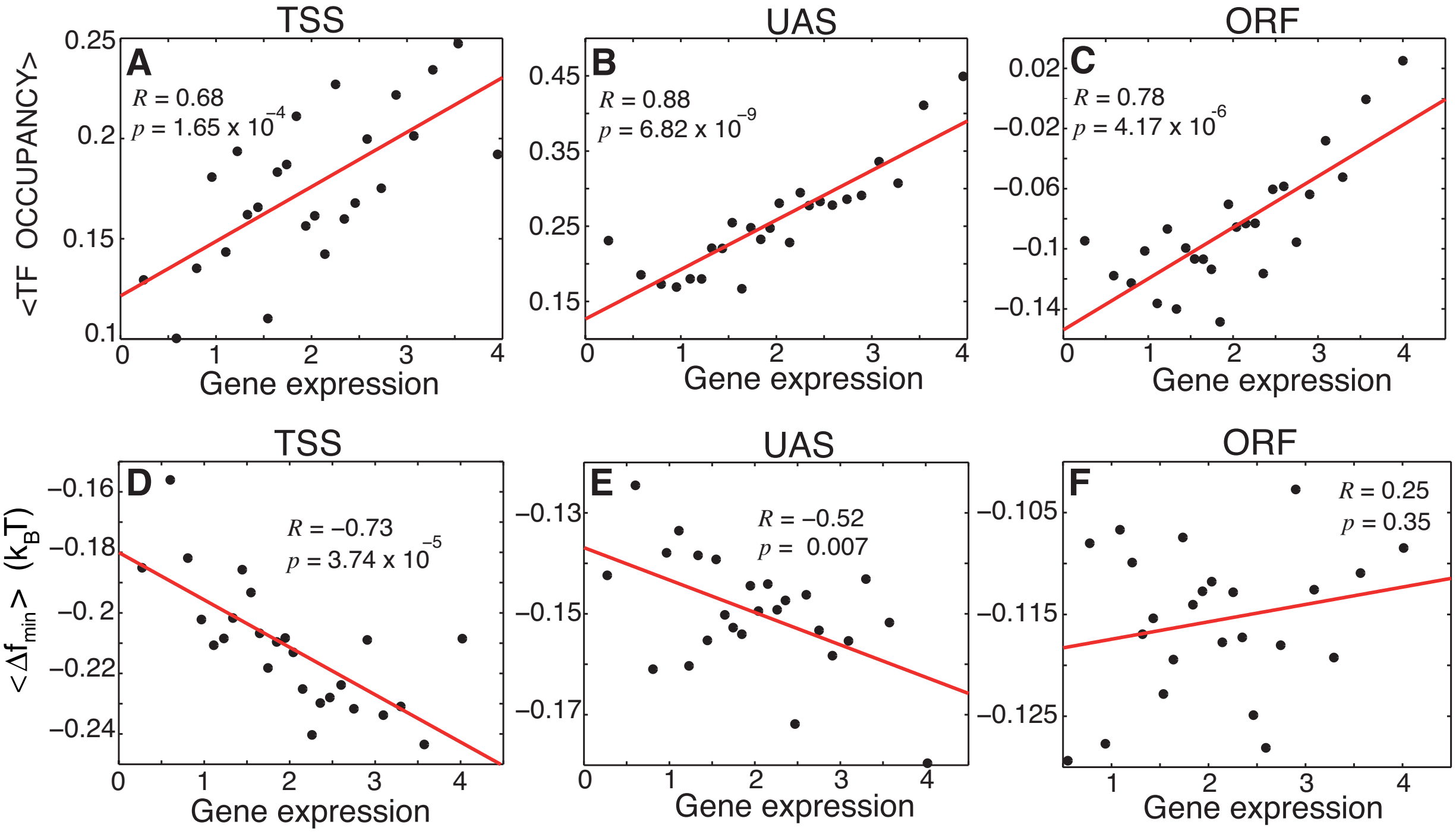

**Figure 5**

## SUPPORTING MATERIAL

## Nonspecific protein-DNA binding is widespread in the yeast genome

Ariel Afek and David B. Lukatsky
*Department of Chemistry, Ben-Gurion University of the Negev, Beer-Sheva 84105 Israel*

## Supporting Figure Legend

**Figure S1.** This figure is analogous to **Figure 1** in the main text, except for the way of how the highest and the lowest average TF occupancy genes are selected in TSS and UAS regions. Here, the experimentally measured average TF occupancy is defined in the following way. First, for each TF the genes are ordered according to their occupancy score for this TF, and 20% of the highest and the lowest occupancy genes are selected for each TF (1). Second, each gene receives a score of how many times it was selected in the highest or the lowest occupancy group. Finally, 10% of the genes with the highest and the lowest scores are selected. Each of these two groups contains 496 genes. **A.** Average free energy of nonspecific TF-DNA binding per bp, $\langle \Delta f \rangle = \langle \Delta F \rangle / M$, computed within the interval (-400,400) for the two groups of genes selected according to the experimentally measured average TF occupancy, as defined above, in the TSS region: 10% highest TF occupancy in the TSS region (red) and 10% lowest TF occupancy in the TSS region (blue). Each group contains 496 genes. Horizontal bar, marked 'TSS', on the *x*-axis, shows the corresponding region where the TF occupancy was measured. **B.** Similar to **(A)**, but the two groups of genes are selected according to the experimentally measured average TF occupancy in the UAS region. Horizontal bar, marked 'UAS', on the *x*-axis, shows the corresponding region where the TF occupancy was measured. The *p*-values where computed in the following way. First, we selected $10^5$ pairs of groups of randomly chosen 496 genes. Second, for each of these pairs of random groups we computed the free energy of nonspecific binding, as described above. Third, within each region of interest (TSS or UAS), we computed the difference between the minima of the average free energy of nonspecific binding, $\langle \Delta f \rangle_{\min}$, for the corresponding pairs of groups. Finally, we computed the probability that this difference is equal or larger than the actual value of the difference.

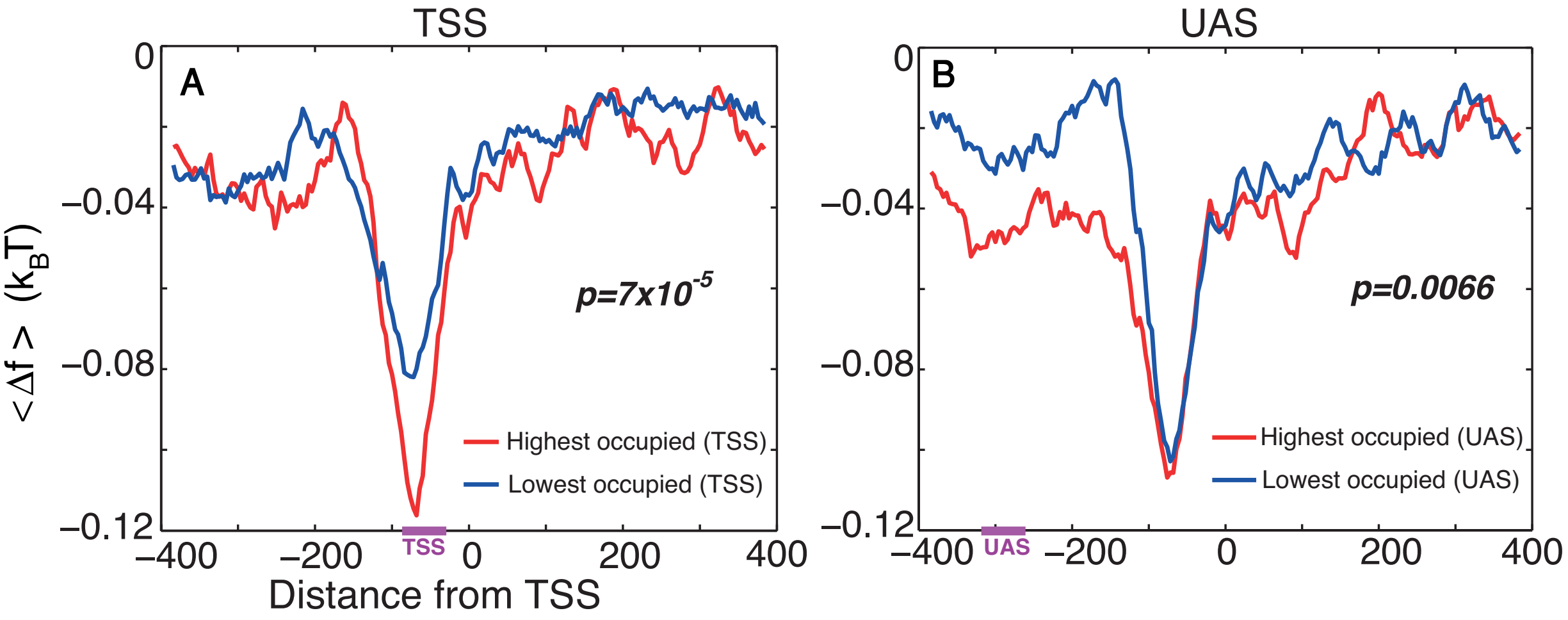

**Figure S1.**